\documentstyle[aps,prd]{revtex}

\begin{document}

\title{Collapsing shells of radiation in anti-de Sitter 
spacetimes and the hoop and cosmic censorship conjectures
\\
gr-qc/9812078
}

\author{Jos\'e P. S. Lemos}

\address{Departamento de Astrof\'{\i}sica,
	      Observat\' orio Nacional-CNPq, Rua General Jos\'e Cristino 77,
	      20921 Rio de Janeiro, Brazil \& \\
         Departamento de F\'{\i}sica,
	      Instituto Superior T\'ecnico, Av. Rovisco Pais 1, 
              1096 Lisboa, Portugal.}

\maketitle

\begin{abstract}
Gravitational collapse of radiation in an anti-de Sitter background 
is studied. For the spherical case, the collapse proceeds in much the 
same way as in the Minkowski background, i.e., massless naked singularities 
may form for a highly inhomogeneous collapse, violating the cosmic 
censorship, but not the hoop conjecture. The toroidal, cylindrical and 
planar collapses can be treated together. In these cases no naked singularity 
ever forms, in accordance with the cosmic censorship. However, since the 
collapse proceeds to form toroidal, cylindrical or planar black holes, 
the hoop conjecture in an anti-de Sitter spacetime is  violated. 

\end{abstract}

\vskip 0.5cm

\noindent
{\bf 1. Introduction}

\vskip 3mm

The cosmic censorship conjecture \cite{penrose69} forbids the existence
of naked singularities, singularities not surrounded by an event
horizon. The hoop conjecture \cite{thorne72} states that BHs form when
and only when a mass $M$ gets compacted into a region whose
circumference in every direction is less than its Schwarzschild
circumference $4\pi M$ ($G=c=1$). The collapse of spherical matter in
the form of dust, or radiation forms massless shell-focusing naked
singularities violating the cosmic censorship 
\cite{eardleysmarr,christodolou,oripiran,
papapetrou,hiscocketal,joshi91,lake92,lemos92}.

It is also violated for matter with cylindrical symmetry where the collapse
proceeds to form a line singularity \cite{apostolatosthorne93}. The
hoop conjecture has not suffered from such counter-examples. 
Cylinders, for instance, do not get compacted, 
and do not form black holes. Spindles form black
holes only when all collapsing directions are sufficiently compactified in
accord with the hoop conjecture \cite{shapiroteukolsky91}. These 
results follow from analysis in asymptotically flat spacetimes. 

In this work we want to study how these features might get modified in
an anti-de Sitter background. We study imploding radiation in an
anti-de Sitter background both in spherically symmetric and toroidally,
cylindrical or plane symmetric spaces, to test the appearance of naked
singularities against the cosmic censorship conjecture and the
formation of toroidal, cylindrical or planar black holes against the
hoop conjecture.

The Vaidya metric, describing a spherically symmetric spacetime with
radiation, is an exact solution of Einstein's field equations which has
been used to generate shell-focusing naked singularities. Its
straightforward generalization to spherically symmetric spacetimes with
negative cosmological constant is called the Vaidya-anti-de Sitter
metric. In addition, to describe collapsing radiation in spacetimes
with toroidal, cylindrical or planar symmetry and a negative
cosmological constant, there is an appropriately modified Vaidya
metric, which of course, satisfies Einstein's field equations.

The spherically symmetric case with zero cosmological constant has been
thoroughly studied (see \cite{joshibook} for a review).  It is usually
admitted that in spherical symmetry the effects of adding a negative
cosmological constant  $\Lambda$ do not alter radically the
description. The main difference is that the exterior spacetime should
be described by the Vaidya-anti-de Sitter metric.  As we shall see the
situation can change drastically for a collapse with non-spherical
topology.  Indeed, we find that (i) in spherical symmetry with negative
$\Lambda$, massless naked singularities form for a sufficiently
inhomogeneous collapse, similarly to the $\Lambda=0$ case and (ii)
toroidal, cylindrical or plane symmetric collapse with negative
$\Lambda$ does not produce naked singularities; instead black holes
will form giving an explicit counter-example to the hoop conjecture.

\vskip .5cm

\noindent
{\bf 2.  Spherical Collapse}

\vskip 3mm

The Einstein field equations are 
\begin{equation}
G_{ab} + \Lambda g_{ab} = 8\pi  T_{ab}, 
                        \label{eq:2.1}
\end{equation}
where $G_{ab}$, $g_{ab}$, $T_{ab}$ are the Einstein, the metric and 
the the energy-momentum tensors, respectively, and $\Lambda$ is the 
cosmological constant ($G=c=1$). In the spherically symmetric case 
the equations admit the Vaidya solution 
\begin{eqnarray}
&ds^2 = - \left(1 + \alpha^2r^2 - \frac{2 m(v)}{ r}\right) dv^2 
+ 2 dv dr +&\nonumber \\
& \quad\quad
+ r^2 (d\theta^2 + \sin^2\theta d\phi^2), &
                         \label{eq:2.2}
\end{eqnarray}
for an energy-momentum tensor given by
\begin{eqnarray}
&T_{ab} = \frac{1}{4\pi r^2} \frac{dm(v)}{dv} k_ak_b, &\nonumber \\
&  \quad k_a = - \delta^v_a, \quad k_ak^a = 0. &
                        \label{eq:2.3}
\end{eqnarray}
Here $\alpha \equiv \sqrt{\frac{-\Lambda}{3}}$, $v$ is a null
coordinate, called the advanced time, with $-\infty<v<\infty$, $r$ is
the radial coordinate with $0<r<\infty$, and $\theta,\phi$ are the
coordinates which describe the two-dimensional spherical surface.  The
Vaidya metric (\ref{eq:2.2}) describes the gravitational field of a
spherical flow of unpolarized radiation in the geometrical optics
approximation. It represents a spherical null fluid.  Noting that the
energy-density of the radiation is $\epsilon = \frac{2}{4\pi 
r^2}\frac{dm}{dv}$, one sees that the weak energy condition for the
radiation is satisfied whenever $\frac{dm}{dv}\geq0$, i.e., the
radiation is imploding. The function $m(v)$ represents a mass and is
thus a non-negative increasing function of $v$.

The physical situation we want to represent is of radiation injected
with the velocity of light radially into an anti-de Sitter spacetime
from infinity towards the center. For $v<0$ the spacetime is anti-de
Sitter with $m(v)=0$. At $v=V$, say, the radiation is turned off. For
$v>V$ the exterior spacetime settles into a Schwarzschild-anti-de
Sitter spacetime. The first spherical ray arrives at the center 
when $r=0$ and $v=0$, forming a singularity. 
We will now test whether future directed null geodesics terminate at the 
singularity. If they do, the singularity is naked.

In these coordinates, lines with $v=$constant represent incoming radial 
null vectors whose generator vectors have the form $k^a=(0,-1,0,0)$, 
with $k_a = (-1,0,0,0)$ (see equation (\ref{eq:2.3})). The generators 
$l^a$ of outgoing null lines are then given by 
$l^a=(1,\frac12(1+\alpha^2r^2-\frac{2m(v)}{ r}),0,0)$
with, $l_al^a=0$ and $l_ak^a=-1$. 
The equation for outgoing radial null geodesics $v(r)$ is then 
\begin{equation}
\frac{dv}{dr} = \frac{l^v}{l^r}= 
\frac{2r}{ r + \alpha^2 r^3 - 2 m(v)}.   
                        \label{eq:2.4}
\end{equation}

Equation (\ref{eq:2.4}), for future null geodesics, has a singular 
point at $r=0$ and $v=0$. The nature of the singular point can now be 
analysed. Following \cite{lake} we write 
\begin{equation}
2m(v) = \lambda v + f(v).  
                        \label{eq:2.5}
\end{equation}
where $\lambda$ is a constant and $f(v)={\rm o}(v)$ as
$v\rightarrow0$.  Following standard techniques \cite{SmithJordan} 
one can see that the singularity is a unstable node when
$0\leq\lambda\leq\frac18$.  Since $\mu\equiv1/\lambda$ gives a degree of
inhomogeniety of the collapse (see \cite{lemos92}), 
we see that for a collapse sufficiently
inhomogeneous naked singularities develop.  This result is the same
result as in collapsing radiation in a Minkowski background, as it
should have been expected, since when $r\rightarrow0$ the cosmological
constant term $\alpha^2 r^2$ is negligible.
The global nakedness of the singularity can then be seen
by making a junction onto the Schwarzschild-anti-de
Sitter spacetime.

The Kretschmann scalar is given by
$K=R^{abcd}R_{abcd}=a\frac{\lambda^2}{r^4}$ for some number $a$. Thus
as $r\rightarrow0$ the collapse forms a polynomial curvature singularity. 
To see if, in addition, the naked singularity is strong we examine the scalar
$\psi\equiv{\rm lim}_{l\rightarrow0}l^2G_{ab}l^al^b$, where $l^a$ is
the null geodesic along the Cauchy horizon parametrized by $l$. This
gives $\psi=4\lambda$, showing that the singularity is strong.

In \cite{joshidwivedi} another solution for outgoing light rays is
presented. This solution is also valid here for sufficiently small radii.
Indeed, it is found that if $2m(v)=\beta
v^\sigma\left(1-2\sigma\beta v^{\sigma-1}\right)$, with $\sigma>1$ and
$\beta>0$ constants, then light rays with $r=v^\sigma$ emanate from a
strong naked singularity.

Thus, for spherical symmetry, the cosmic censorship is violated for a
sufficiently inhomogeneous collapse. On the other hand, the hoop
conjecture is validated since a black hole forms whenever the spherical
surface of matter passes its own gravitational radius.

\vskip .5cm

\noindent
{\bf 3.  Toroidal, Cylindrical or Planar Collapse}

\vskip 3mm

We now study the phenomenon of gravitational collapse of toroidal
cylindrical or planar shells of radiation to verify whether naked
singularities can occur.

Einstein's filed equations (\ref{eq:2.1}), have also the solution 
\begin{equation}
ds^2 = - \left(\alpha^2r^2-\frac{q m(v)}{r}\right) dv^2 + 2 dv dr + 
r^2 (d\theta^2 + d\phi^2),
                         \label{eq:3.1}
\end{equation}
for an energy-momentum tensor given by
\begin{eqnarray}
&T_{ab} = \frac{q}{8\pi r^2} \frac{dm(v)}{dv} k_ak_b, &\nonumber \\
&  \quad k_a = - \delta^v_a, \quad k_ak^a = 0, &
                        \label{eq:3.2}
\end{eqnarray}
where now, $\theta,\phi$ are the coordinates which describe the
two-dimensional zero-curvature space generated by the two-dimensional
commutative Lie group $G_2$ of isometries. The topologies of this
two-dimensional space can be (i) $T^2=S^1 \times S^1$, the flat torus
model $[G_2=U(1)\times U(1)]$, (ii) $R\times S^1$, the cylindricallly
symmetric model $[G_2=R\times U(1)]$, and (iii) $R^2$, the planar model
$[G_2=E_2]$.  In the toroidal case we choose $0\leq\theta<2\pi$,
$0\leq\phi<2\pi$, in the cylindrical case $-\infty<\theta<\infty$,
$0\leq\phi<2\pi$, and in the planar case $-\infty<\theta<\infty$,
$-\infty<\phi<\infty$. The parameter $q$ has different values depending
on the topology of the two-dimensional space. For the torus
$q=\frac{2}{\pi}$ and $m(v)$ is a mass, for the cylinder
$q=\frac{4}{\alpha}$ and $m(v)$ is a mass per unit length, and for the
plane $q=\frac{2}{\alpha^2}$ and $m(v)$ is a mass per unit area. The
values of the parameter $q$ given above, were taken from ADM masses of
the corresponding static BHs found in
\cite{lemoszanchinprd96,lemosplb95,lemoscqg95}. 
Setting $m(v)=$const and $\alpha=0$ one obtains 
the Taub metric \cite{taub}. In this case, when $\alpha=0$, there are 
no black hole solutions \cite{lemoscqg95}.

As in the spherical symmetric case, 
metric (\ref{eq:3.1})  describes the gravitational 
field of a toroidal, cylindrical or planar flow of unpolarized 
radiation in the geometrical optics approximation. 
In the same way, for imploding radiation, $\frac{dm}{dv}\geq0$, 
one has $\epsilon = \frac{q}{8\pi r^2}\frac{dm}{dv}>0$, 
and the mass $m(v)$ is a non-negative increasing function of $v$.

One now injects radiation radially towards the center. 
If one whishes, one turns off the radiation at $v=V$, say, 
and then makes a junction with the exterior background spacetime, 
sometimes called Riemann-anti de Sitter spacetime \cite{vanzo}. 
The first spherical ray arrives at the center 
for $r=0$, $v=0$, forming a singularity.

Lines with $v=$constant represent incoming radial 
null geodesics, with generators $k^a=(0,-1,0,0)$, 
The generators 
$l^a$ of outgoing null lines are
$l^a=(1,\frac12(\alpha^2r^2-\frac{qm(v)}{r}),0,0)$. 
The equation for outgoing radial null geodesics $v(r)$ is now
\begin{equation}
\frac{dv}{dr} = \frac{l^v}{l^r}= 
 \frac{2 r}{\alpha^2 r^3 - q m(v)}.  
                        \label{eq:3.3}
\end{equation}
Again, equation (\ref{eq:3.3}) for future null geodesics has a singular 
point at $r=0$ and $v=0$, and we write as in (\ref{eq:2.5})
\begin{equation}
qm(v) = \lambda v + {\rm o}(v).  
                        \label{eq:3.4}
\end{equation}
Following standard techniques one finds that the singularity is a
center for all $\lambda$, and no radial future null geodesics terminate
at the singularity. This is in contrast with the spherically symmetric
collapse. In turn the collapse proceeds to form toroidal, cylindrical
or planar black holes, once the exterior event horizon condition is
achieved, i.e., $r=\left(\frac{qM}{\alpha^2}\right)^{1/3}$, where 
$M$ is the total mass of the collapsing radiation, (see also 
\cite{lemostorcol}).

One can also try the Joshi and Dwivedi's  type solutions
\cite{joshidwivedi}. However these give for the mass $q 
m(v)=\beta \alpha^2 v^3 - 2\beta^{\frac23}  v$, for some
arbitrary constant $\beta$.  They are unphysical since
$\frac{dm}{dv}|_{v=0} <0$.

Thus, for toroidal, cylindrical or planar collapse of radiation, naked
singularities do not form, in accord with the cosmic censorship 
conjecture. On the other hand, cylindrical black holes as well as 
toroidal and planar ones do form, giving an explicit example of 
violation of the hoop conjecture. 

Note that these conclusions are valid as long as $\alpha>0$. For 
$\alpha=0$ the collapse proceeds to form a naked singularity with 
toroidal, cylindrical, or planar symmetry and the Taub solution 
is the final background spacetime. In such case the cosmic censorship 
is violated, whereas the hoop conjecture is not 
\cite{thorne72,apostolatosthorne93}.

\vskip .5cm

\noindent
{\bf 4. Conclusions}

\vskip 3mm

The Vaidya metric has been extensively used to study the formation of naked
singularities in spherical gravitational collapse \cite{joshibook}. We
have extended this study here to include a negative cosmological
constant, and found that locally, in the vicinity of the singularity,
the same results prevail.
On the other hand, we have found that collapse of toroidal, cylindrical
or planar radiation in a spacetime with negative cosmological constant 
does not yield naked singularities, but always 
black holes. This shows that non-spherical collapse in a negative
cosmological constant background may violate the hoop but not the
cosmic censorship conjecture.

\end{document}